\documentclass[runningheads,a4paper]{llncs}
\usepackage{amssymb}
\usepackage{amsmath}
\usepackage{graphicx}
\usepackage{url}

\newcommand{\keywords}[1]{\par\addvspace\baselineskip\noindent\keywordname\enspace\ignorespaces#1}

\begin{document}
\mainmatter  

\title{The Directed Dominating Set Problem: Generalized Leaf Removal and Belief Propagation}
\titlerunning{Directed Dominating Set Problem}

\author{
  Yusupjan Habibulla \and
  Jin-Hua Zhao \and
  Hai-Jun Zhou
}
\authorrunning{Y. Habibulla, J.-H. Zhao, and H.-J. Zhou}

\institute{
  State Key Laboratory of Theoretical Physics, Institute of Theoretical 
  Physics,\\
  Chinese Academy of Sciences,  Beijing 100190,  China
}

\toctitle{Lecture Notes in Computer Science}
\maketitle

\begin{abstract} 
  A minimum dominating set for a digraph (directed graph) is a smallest set of
  vertices such that each vertex either belongs to this set or has at least one
  parent vertex in this set. We solve this hard combinatorial optimization
  problem approximately by a local algorithm of generalized leaf removal and by
  a message-passing algorithm of belief propagation. These algorithms can
  construct near-optimal dominating sets or even exact minimum dominating sets
  for random digraphs and also for real-world digraph instances. We further
  develop a core percolation theory and a replica-symmetric spin glass theory
  for this problem.  Our algorithmic and theoretical results may facilitate
  applications of dominating sets to various network problems involving
  directed interactions.
  
\end{abstract}

\section{Introduction}

The construction of a minimum dominating set (MDS) for a general digraph
(directed graph) \cite{Fu-1968,Haynes-Hedetniemi-Slater-1998} is a fundamental
nondeterministic polynomial-hard (NP-hard) combinatorial optimization problem 
\cite{Garey-Johnson-1979}. A digraph $D = \{V, A\}$ is formed by a set 
$V \equiv \{1, 2,.., N \}$ of $N$ vertices and a set
$A\equiv \{(i, j) : i, j \in V\}$ of $M$ arcs (directed edges), each arc
$(i, j)$ pointing from a parent vertex (predecessor) $i$ to a child vertex
(successor) $j$. The arc density $\alpha$ is defined simply as
$\alpha \equiv M / N$. Each vertex $i$ of digraph $D$ brings a constraint
requiring that either $i$ belongs to a vertex set $\Gamma$ or at least one of
its predecessors belongs to $\Gamma$. A dominating set $\Gamma$ is therefore a
vertex set which satisfies all the $N$ vertex constraints, and the dominating
set problem can be regarded as a special case of the more general hitting set
problem \cite{Mezard-Tarzia-2007,Gutin-Jones-Yeo-2011}.

A dominating set containing the smallest number of vertices is a MDS, which
might not necessarily be unique for a digraph $D$. As a MDS is a smallest set
of vertices which has directed edges to all the other vertices of a given
digraph, it is conceptually and practically important for analyzing, monitoring,
and controlling many directed interaction processes in complex networked
systems, such as infectious disease spreading
\cite{Takaguchi-Hasegawa-Yoshida-2014}, genetic regulation
\cite{Wuchty-2014,Wang-etal-2014}, chemical reaction and metabolic regulation
\cite{Liu-Slotine-Barabasi-2013}, and power generation and transportation 
\cite{Yang-Wang-Motter-2012}. Previous heuristic algorithms on the directed MDS
problem all came from the computer science/applied mathematics communities 
\cite{Haynes-Hedetniemi-Slater-1998} and they are based on vertices' local
properties such as in- and out-degrees 
\cite{Pang-etal-2010,Takaguchi-Hasegawa-Yoshida-2014,Molnar-etal-2013}. In the
present work we study  the directed MDS problem through statistical mechanical
approaches. 

In the next section we introduce a generalized leaf-removal (GLR) process to 
simplify an input digraph $D$. If GLR reduces the original digraph $D$ into an
empty one, it then succeeds in constructing an exact MDS. If a core is left
behind, we implement a hybrid algorithm combining GLR with an impact-based
greedy process to search for near-optimal dominating sets (see 
Fig.~\ref{fig:mdsd_er_rr} and Table~\ref{tab:real}). We also study the
GLR-induced core percolation by a mean field theory (see
Fig.~\ref{fig:glrd_er_rr}). In Sec.~\ref{sec:bp} we introduce a spin glass
model for the directed MDS problem and obtain a belief-propagation decimation
(BPD) algorithm based on the replica-symmetric mean field theory. By comparing
with ensemble-averaged theoretical results, we demonstrate that the
message-passing BPD algorithm has excellent performance on random digraphs and
real-world network instances, and it outperforms the local hybrid algorithm
(Fig.~\ref{fig:mdsd_er_rr} and Table~\ref{tab:real}). 

This paper is a continuation of our earlier effort
\cite{Zhao-Habibulla-Zhou-2014} which studied the undirected MDS problem.
Since each undirected edge between two vertices $i$ and $j$ can be treated as
two opposite-direction arcs $(i,j)$ and $(j, i)$, the methods of this paper
are more general and they are applicable to graphs with both directed and
undirected edges. The algorithmic and theoretical results presented here and
in \cite{Zhao-Habibulla-Zhou-2014} may promote the application of dominating
sets to various network problems involving directed and undirected interactions.

In the remainder of this paper, we denote by $\partial i^{+}$ the set of
predecessors of a vertex $i$, and refer to the size of this set as the
in-degree of $i$; similarly $\partial i^{-}$ denotes the set of successors of
vertex $i$ and its size defines the out-degree of this vertex. With respective
to a dominating set $\Gamma$, if vertex $i$ belongs to this set, we say $i$ is
occupied, otherwise it is unoccupied (empty). If vertex $i$ belongs to the
dominating set $\Gamma$ or at least one of its predecessors belongs to $\Gamma$,
then we say $i$ is observed, otherwise it is unobserved.

\section{Generalized Leaf Removal and the Hybrid Algorithm}
\label{sec:hybrid}

\begin{figure}[t]
  \begin{center}
    \includegraphics[width=0.6\linewidth]{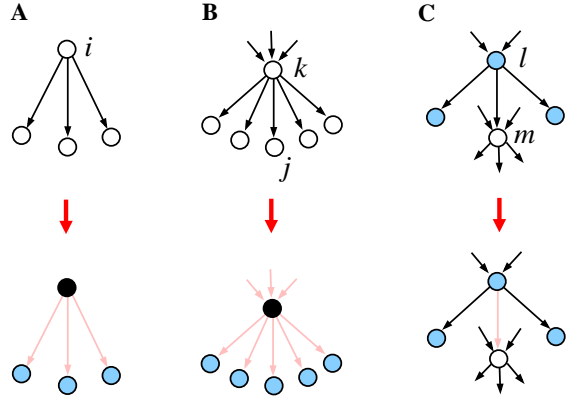}
  \end{center}
\caption{\label{fig:glr}
  The generalized leaf-removal process. White circles represent unobserved
  vertices, black circles are occupied vertices, and blue (gray) circles are
  observed but unoccupied vertices. Pink (light gray) arrows represent deleted
  arcs, while black arrows are arcs that are still present in the digraph. (A)
  vertex $i$ has no predecessor, so it is occupied. (B) vertex $j$ has only one
  predecessor $k$ and no successor, so vertex $k$ is occupied. (C) vertex $l$
  has only a single unobserved successor $m$, so the arc $(l, m)$ is deleted.}
\end{figure}

The leaf-removal process was initially applied in the vertex-cover problem
\cite{Bauer-Golinelli-2001}. It causes a core percolation phase transition in
random undirected or directed graphs \cite{Liu-etal-2012}. Here we consider a
generalized leaf-removal process for the directed MDS problem. This GLR process
iteratively deletes vertices and arcs from an input digraph $D$ starting from
all the $N$ vertices being unoccupied (and unobserved) and the dominating set
$\Gamma$ being empty. The microscopic rules of digraph simplification are as
follows:

Rule $1$: If an unobserved vertex $i$ has no predecessor in the current digraph
$D$, it is added to set $\Gamma$ and become occupied (see Fig.~\ref{fig:glr}A).
All the previously unobserved successors of $i$ then become observed.

Rule $2$: If an unobserved vertex $j$ has only a single unoccupied predecessor
(say vertex $k$) and no unobserved successor in the current digraph $D$, vertex
$k$ is added to set $\Gamma$ and become occupied (Fig.~\ref{fig:glr}B). All the
previously unobserved successors of $k$ (including $j$) then become observed.

Rule $3$: If an unoccupied but observed vertex $l$ has only a single unobserved
successor (say $m$) in the current digraph $D$, occupying $l$ is \emph{not}
better than occupying $m$, therefore the arc $(l, m)$ is deleted from $D$
(Fig.~\ref{fig:glr}C). We emphasize that vertex $m$ is still unobserved after
this arc deletion. (Rule $3$ is specific to the dominating set problem and it
is absent in the conventional leaf-removal process
\cite{Bauer-Golinelli-2001,Liu-etal-2012}.)

The above-mentioned microscopic rules only involve the local structure of the
digraph, they are simple to implement. Following the same line of reasoning in 
\cite{Zhao-Habibulla-Zhou-2014}, we can prove that if all the vertices are
observed after the GLR process, the constructed vertex set $\Gamma$ must be a
MDS for the original digraph $D$. If some vertices remain to be unobserved after
the GLR process, this set of remaining vertices is unique and is independent of
the particular order of the GLR process.

\subsection{Core percolation transition}

\begin{figure}[t]
  \begin{center}
    \includegraphics[angle=270, width = 1.0\linewidth]{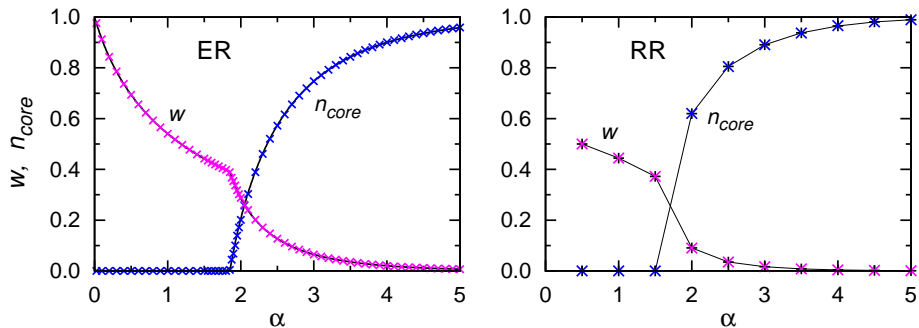}
  \end{center}
  \caption{
    \label{fig:glrd_er_rr}
    GLR-induced core percolation transition in Erd\"{o}s-R\'{e}nyi (left panel)
    and regular random (right panel) digraphs. $w$ is the fraction of occupied
    vertices, $n_{core}$ is the fraction of remaining unobserved vertices. Cross
    symbols are results obtained on a single digraph with $N=10^6$ vertices and
    $M = \alpha N$ arcs, lines (left panel) and plus symbols connected by lines 
    (right panel) are mean-field theoretical results for $N=\infty$. }
\end{figure}

We apply GLR  on a set of random Erd\"{o}s-R\'{e}nyi (ER) digraphs and random
regular (RR) digraphs (see Fig.~\ref{fig:glrd_er_rr}) and also on a set of
real-world directed networks (see Table~\ref{tab:real}). To generate an ER
digraph of size $N$ and arc density $\alpha$, we first select $\alpha N$
different pairs of vertices totally at random from the set of $N (N-1)/2$
possible pairs, and then create an arc of random  direction between each
selected vertex pair. Similarly, to generate a RR digraph, we first generate an
undirected RR graph with every vertex having the same integer number
($=2 \alpha$) of edges \cite{Zhao-Habibulla-Zhou-2014}, and then randomly
specify a direction for each undirected edge.

If the arc density $\alpha$ of an ER digraph is less than $1.852$ and that of a
RR digraph is less than $2.0$, a MDS can be constructed by applying GLR alone.
However, if $\alpha > 1.852$ for an ER digraph and $\alpha \geq 2.0$ for a RR
digraph, GLR only constructs a partial dominating set for the digraph, and a
fraction $n_{core}$ of vertices remain to be unobserved after the termination of
GLR. For ER digraphs $n_{core}$ increases continuously from zero as $\alpha$
exceeds $1.852$. The sub-digraph induced by all these unobserved vertices and
all their predecessor vertices is referred to as the core of digraph $D$.

We develop a percolation theory to quantitatively understand the GLR dynamics
on random digraphs. For theoretical simplicity we consider a GLR process
carried out in discrete time steps $t=0, 1, \ldots$. In each time step $t$,
first Rule $1$ is applied to all the eligible vertices, then Rule $2$ is
applied to all the eligible vertices, then Rule $3$ is applied to all the
eligible arcs, and finally all the newly occupied vertices and their attached
arcs are all deleted from digraph $D$. The fraction $w$ of occupied vertices
during the whole GLR process and the fraction $n_{core}$ of remaining unobserved
vertices are quantitatively predicted by this mean-field theory (see the
Appendix for technical details). These theoretical predictions are in complete
agreement with simulation results on single digraph instances
(Fig.~\ref{fig:glrd_er_rr}). We believe that when there is no core
($n_{core}=0$), the MDS relative size $w$ as predicted by our theory is the
exact ensemble-averaged result for finite-connectivity random digraphs.

\subsection{The hybrid algorithm}

The GLR process can not construct a MDS for the whole digraph $D$ if it
contains a core. For such a difficult case we combine GLR with a simple
greedy process to construct a dominating set that is not necessarily a MDS.
We define the impact of an unoccupied vertex as the number of newly observed
vertices caused by occupying this vertex
\cite{Haynes-Hedetniemi-Slater-1998,Takaguchi-Hasegawa-Yoshida-2014,Molnar-etal-2013}.
For example, an unobserved vertex with three unobserved successors has 
impact $4$, while an observed vertex with three unobserved successors has 
impact $3$.
Our hybrid algorithm has two modes, the default mode and the greedy mode.
In the default mode, the digraph is iteratively simplified by occupying
vertices according to the microscopic rules
of GLR. If there are still unobserved vertices after this process,
the algorithm first switches to the greedy mode, in which the digraph is 
simplified by occupying a vertex randomly chosen from the subset of 
highest-impact vertices, and then switches back to the default mode.

\begin{table}[t]
  \caption{
    \label{tab:real}
    Constructing dominating sets for several real-world network instances
    containing $N$ vertices and $M = \alpha N$ arcs. For each graph, we list
    the number of unobserved vertices after the GLR process (Core), the size of
    the dominating set obtained by a single running of the greedy algorithm
    (Greedy), the hybrid algorithm (Hybrid), and the BPD algorithm at fixed
    re-weighting parameter $x=8.0$ (BPD). Epinions1
    \cite{Richardson-Agrawal-Domingos-2003} and WikiVote 
    \cite{Leskovec-Huttenlocher-Kleinberg-2010,Leskovec-Huttenlocher-Kleinberg-2010-b}
    are two social networks, Email \cite{Leskovec-Kleinberg-Faloutsos-2007} and
    WikiTalk 
    \cite{Leskovec-Huttenlocher-Kleinberg-2010,Leskovec-Huttenlocher-Kleinberg-2010-b}
    are two communication networks, HepPh and HepTh \cite{Leskovec-etal-2005}
    are two research citation networks, Google and Stanford
    \cite{Leskovec-etal-2009} are two webpage connection networks, and
    Gnutella31 \cite{Ripeanu-etal-2002} is a peer-to-peer network. }
  \begin{center}
    \begin{tabular}{rrrrrrrrr}
      \hline
      Network    & $N$       & $M$       &  $\alpha$  & Core    
      & Greedy   & Hybrid    & BPD \\ \hline
      Epinions1  & $75879$   & $405740$  &  $5.347$   & $348$  
      & $37172$  & $37128$   & $37127$ \\
      WikiVote   & $7115$    & $100762$  &  $14.162$  & $7$    
      & $4786$   & $4784$    & $4784$ \\
      Email      & $265214$  & $364481$  &  $1.374$   & $0$   
      & \quad $203980$ & $203980$  & $203980$ \\
      WikiTalk   & \quad $2394385$ & $4659565$ &  $1.946$   & $72$  
      & $63617$  & $63614$   & $63614$ \\
      HepPh      & $34546$   & $420877$  &  $12.183$  & $982$  
      & $9628$   & $9518$    & $9512$ \\
      HepTh      & $27770$   & $352285$  & \quad $12.686$  & $1900$ 
      & $7302$   & $7213$    & $7203$    \\
      Google     & $875713$  & \quad $4322051$ &  $4.935$   & $98473$
      &  $315585$ & \quad $314201$  & \quad $313986$ \\
      Stanford   & $281903$  & $1992636$ &  $7.069$   & \quad $68947$ 
      & $90403$  & $89388$   & $89466$     \\
      Gnutella31 & $62586$   & $147892$  &  $2.363$   & $26$  
      & $12939$  & $12784$   & $12784$ \\
      \hline
    \end{tabular}
  \end{center}
\end{table}

The hybrid algorithm can be regarded as an extension of the pure greedy
algorithm which always works in the greedy mode.
The simulation results obtained by the hybrid algorithm and the pure
greedy algorithm are shown in Fig.~\ref{fig:mdsd_er_rr} for random
digraphs and in Table~\ref{tab:real} for real-world network instances.
The hybrid algorithm improves over the greedy algorithm considerably on
random digraph instances when the arc density $\alpha \leq 10$. 
But when the relative size $n_{core}$ of the core in
the digraph is close to $1$, the hybrid algorithm only slightly outperforms
the pure greedy algorithm.

\section{Spin Glass Model and Belief-Propagation}
\label{sec:bp}

We now introduce a spin glass model for the directed MDS problem and solve it
by the replica-symmetric mean field theory, which is based on the Bethe-Peierls
approximation \cite{Mezard-Montanari-2009,Kschischang-etal-2001} but can also
be derived without any physical assumptions through partition function expansion
\cite{Xiao-Zhou-2011,Zhou-Wang-2012}. We define a partition function $Z(x)$ for
a given input digraph $D$ as follows:
\begin{equation}
  \label{eq:Z}
  Z(x)  = \sum\limits_{\underline c}  \prod\limits_{i \in V} \Bigl[ e^{- x  c_i}
    \bigl(1 -(1-c_i) \prod\limits_{j \in \partial i^{+}} (1-c_j) \bigr) \Bigr] \; .
\end{equation}
The summation in this expression is over all the microscopic configurations 
$\underline{c} \equiv \{c_1, c_2, ... , c_N\}$ of the $N$ vertices, with 
$c_i \in \{0, 1\}$ being the state of vertex $i$ ($c_i=0$, empty; $c_i=1$, 
occupied). A configuration $\underline{c}$ has zero contribution to $Z(x)$
if it does not satisfy all the vertex constraints; if it does satisfy all
these constraints and therefore is equivalent to a dominating set, it
contributes a statistical weight $e^{- x W(\underline{c})}$, with 
$W(\underline{c}) \equiv \sum_{i\in V} c_i$ being the total number of occupied
vertices. When the positive re-weighting parameter $x$ is sufficiently large, 
$Z(x)$ will be overwhelmingly contributed by the MDS configurations.

\begin{figure}[t]
  \begin{center}
    \includegraphics[angle=270,width = 1.0\linewidth]{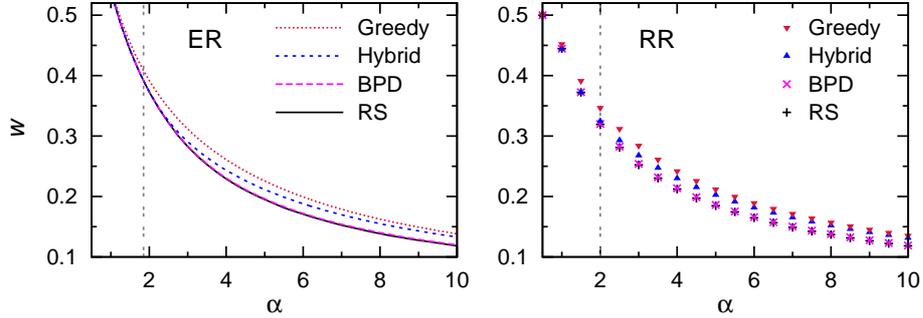}
  \end{center}
  \caption{
    \label{fig:mdsd_er_rr}
    Relative sizes $w$ of dominating sets for Erd\"{o}s-R\'{e}nyi (left panel)
    and random regular (right panel) digraphs. We compare the mean sizes of
    $96$ dominating sets obtained by the Greedy, the Hybrid, and the BPD
    algorithm on $96$ digraph instances of size $N=10^5$ and arc density
    $\alpha$ (fluctuations to the mean are of order $10^{-4}$ and are not
    shown). The MDS relative sizes predicted by the replica-symmetric theory
    are also shown. The re-weighting parameter is fixed to $x=10.0$ for ER
    digraphs and to $x=8.0$ for RR digraphs. The vertical dashed lines mark the
    core-percolation transition point $\alpha \approx 1.852$ for ER digraphs
    and $\alpha=2.0$ for RR digraphs. }
\end{figure}

We define on each arc $(i, j)$ of digraph $D$ a distribution function
$q_{i \rightarrow j}^{c_i, c_j}$, which is the probability of vertex $i$ being in
state $c_i$ and vertex $j$ being in state $c_j$ if all the other attached arcs
of $j$ are deleted and the constraint of $j$ is relaxed, and another
distribution function $q_{j \leftarrow  i}^{c_j, c_i}$, which is the probability of
$i$ being in state $c_i$ and $j$ being in state $c_j$ if all the other attached
arcs of $i$ are deleted and the constraint of $i$ is relaxed. Assuming all the
neighboring vertices of any vertex $i$ are mutually independent of each other
when the constraint of vertex $i$ is relaxed (the Bethe-Peierls approximation),
then when this constraint is present, the marginal probability $q_i^{c_i}$ of
vertex $i$ being in state $c_i$ is estimated by
\begin{equation}
  \label{eq:bp_qi}
  q_{i}^{c_i} = \frac{1}{z_i} e^{- x c_i} \Bigl[\prod\limits_{j \in \partial i^{+}}
    \sum\limits_{c_j} q_{j \rightarrow i}^{c_j, c_i} - \delta_{0}^{c_i}
    \prod\limits_{j \in \partial i^{+}} q_{j \rightarrow i}^{0, 0} \Bigr]
  \prod\limits_{k\in \partial i^{-}} \sum\limits_{c_k} q_{k \leftarrow i}^{c_k, c_i} \; ,
\end{equation}
where $z_i$ is a normalization constant, and $\delta_{m}^{n}$ is the Kronecker
symbol with $\delta_{m}^{n}=1$ if $m=n$ and $\delta_{m}^{n}=0$ if otherwise.
Under the same approximation we can derive the following Belief-Propagation
(BP) equations on each arc $(i, j)$:
\begin{subequations}
  \label{eq:bp}
  \begin{align}
    q_{i \rightarrow j}^{c_i, c_j} & = \frac{1}{z_{i\rightarrow j}} e^{- x c_i}
    \Bigl[\prod\limits_{k \in \partial i^{+}} \sum\limits_{c_k}
      q_{k \rightarrow i}^{c_k, c_i} - \delta_{0}^{c_i} \prod\limits_{k \in \partial i^{+}} 
      q_{k \rightarrow i}^{0, 0} \Bigr] \prod\limits_{l \in \partial i^{-} \backslash j} 
    \sum\limits_{c_l} q_{l \leftarrow i}^{c_l, c_i} \; ,  \label{eq:bp_qij} \\
    q_{j \leftarrow i}^{c_j, c_i}  &  = \frac{1}{z_{j\leftarrow i}} e^{- x c_j}
    \Bigl[ \prod\limits_{k \in \partial j^{+} \backslash i} \sum\limits_{c_k}
      q_{k \rightarrow j}^{c_k, c_j} - \delta_{0}^{c_j + c_i} \prod\limits_{k \in \partial 
        j^{+} \backslash i} q_{k \rightarrow j}^{0, 0} \Bigr]
    \prod\limits_{l\in \partial j^{-}} \sum\limits_{c_l} q_{l \leftarrow j}^{c_l, c_j} \; ,
    \label{eq:bp_qji}
  \end{align}
\end{subequations}
where $z_{i\rightarrow j}$ and $z_{j\leftarrow i}$ are also normalization constants,
and $\partial j^{+} \backslash i$ is the vertex set obtained after removing $i$
from $\partial j^{+}$. We can easily verify that
$q_{i \rightarrow j}^{c_i, 0} = q_{i \rightarrow j}^{c_i, 1}$ for $c_i=0$ or $1$, and
that $q_{j\leftarrow  i}^{1, 0} = q_{j \leftarrow  i}^{1, 1}$.

We let Eqs.~(\ref{eq:bp_qi}) and (\ref{eq:bp}) guide our construction of a
near-optimal dominating set $\Gamma$ through a belief propagation decimation
algorithm. This BPD algorithm is implemented in the same way as the BPD
algorithm for undirected graphs \cite{Zhao-Habibulla-Zhou-2014}, therefore its
implementing details are omitted here (the source code is available upon
request). Roughly speaking, at each iteration step of BPD we first iterate
Eq.~(\ref{eq:bp}) for several rounds, then we estimate the occupation
probabilities for all the unoccupied vertices using Eq.~(\ref{eq:bp_qi}), and
then we occupy those vertices whose estimated occupation probabilities are the
highest. Such a BPD process is repeated on the input digraph until all the
vertices are observed. The results of this message-passing algorithm are shown
in Fig.~\ref{fig:mdsd_er_rr} for random digraphs and in Table~\ref{tab:real}
for real-world networks.

If we can find a fixed point for the set of BP equations at a given value of
the re-weighting parameter $x$, we can then compute the mean fraction $w$ of
occupied vertices as $w = (1/N) \sum_{i\in V} q_i^{1}$. The total free energy
$F= -(1/x) \ln Z(x)$ can be evaluated as the total vertex contributions
subtracting the total arc contributions:
\begin{eqnarray}
  F & = & -\sum\limits_{i\in V} \frac{1}{x} \ln \biggl[\sum\limits_{c_i}
    e^{- x c_i} \Bigl[ \prod\limits_{j \in \partial i^{+}} \sum\limits_{c_j}
      q_{j \rightarrow i}^{c_j, c_i} - \delta_{0}^{c_i} \prod\limits_{j \in \partial i^{+}} 
      q_{j \rightarrow i}^{0, 0} \Bigr] \prod\limits_{k \in \partial i^{-}} 
    \sum\limits_{c_k} q_{k \leftarrow i}^{c_k, c_i} \biggr] \nonumber \\
  & & + \sum\limits_{(i,j)\in A} \frac {1}{x} \ln \Bigl[\sum\limits_{c_i, c_j}
    q_{i \rightarrow j}^{c_i, c_j} q_{j \leftarrow i}^{c_j, c_i} \Bigr] \; .
  \label{eq:free}
\end{eqnarray}
The entropy density $s$ of the system is then estimated through $s = x(w-F/N)$.

For a given ensemble of random digraphs, the ensemble-averaged occupation
fraction $w$ and entropy density $s$ at each fixed value of $x$ can also be
obtained from Eqs.~(\ref{eq:bp_qi}), (\ref{eq:bp}) and (\ref{eq:free}) through
population dynamics simulation \cite{Zhao-Habibulla-Zhou-2014}. Both $w$ and
$s$ decrease with $x$, and $s$ may change to be negative as $x$ exceeds certain
critical value. The value of $w$ at this critical point of $x$ is then taken as
the ensemble-averaged MDS relative size $w_0$ (very likely it is only a lower
bound to $w_0$). For example, at arc density $\alpha=5$ the entropy density of
ER digraphs decreases to zero at $x\approx 9.9$, at which point
$w\approx 0.195$. These ensemble-averaged results for random ER and RR digraphs
are also shown in Fig.~\ref{fig:mdsd_er_rr}. We notice that the BPD results and
the replica-symmetric mean field results almost superimpose with each other,
suggesting that dominating sets obtained by the BPD algorithm are extremely
close to be optimal.

\section{Conclusion}
\label{sec:dis}

In this paper we studied the directed dominating set problem by a
core percolation theory and a replica-symmetric mean field theory, and
proposed a generalized leaf-removal local algorithm and a BPD message-passing
algorithm to construct near-optimal dominating sets for single digraph
instances. 
We expect these theoretical and algorithmic results to be useful
for many future practical applications. 

The spin glass model (\ref{eq:Z}) was treated in this paper only at the
replica-symmetric mean field level. It should be interesting to extend the
theoretical investigations to the level of replica-symmetry-breaking
\cite{Mezard-Parisi-2001} for a more complete understanding of this spin glass
system. The replica-symmetry-breaking mean field theory can also lead to other
message-passing algorithms that perform even better than the BPD algorithm
\cite{Mezard-Montanari-2009} (the review paper \cite{Zhao-Zhou-2014} offers a
demonstration of this point for the minimum vertex-cover problem).

\subsubsection*{Acknowledgments.}
This research is partially supported by the National Basic Research Program
of China (grant number 2013CB932804) and by the National Natural Science
Foundations of China (grant numbers 11121403 and 11225526). HJZ conceived
research, JHZ and YH performed research, HJZ and JHZ wrote the paper.
Correspondence should be addressed to HJZ ({\tt zhouhj@itp.ac.cn}) or to JHZ
({\tt zhaojh@itp.ac.cn}).



\appendix

\section*{Appendix: Mean field equations for the GLR process}
\label{app:glr}

The mean field theory for the directed GLR process is a simple 
extension of the same 
theory presented in \cite{Zhao-Habibulla-Zhou-2014} for undirected graphs.
Therefore here we only list the  main equations of this theory
but do not give the derivation details.
We denote by $P(k_{+}, k_{-})$ the probability that a randomly chosen
vertex of a digraph has in-degree $k_{+}$ and out-degree $k_{-}$.
Similarly, the in- and out-degree joint probabilities
of the predecessor vertex $i$ and successor vertex $j$ of
a randomly chosen arc $(i, j)$ of
the digraph are denoted as $Q_{+}(k_{+}, k_{-})$ and
$Q_{-}(k_{+}, k_{-})$, respectively. We assume that there is no structural
correlation in the digraph, therefore 
\begin{equation}
  Q_{+}(k_{+}, k_{-}) = \frac{ k_{-} P(k_{+}, k_{-})}{\alpha} \; , \quad
  Q_{-}(k_{+}, k_{-}) = \frac{ k_{+} P(k_{+}, k_{-})}{\alpha} \; ,
\end{equation}
where 
$\alpha \equiv \sum_{k_{+}, \; k_{-}} k_{+} P(k_{+}, k_{-}) =
\sum_{k_{+}, \; k_{-}} k_{-} P(k_{+}, k_{-})$ is the arc density.

Consider a randomly chosen arc $(i, j)$ from vertex $i$ to vertex $j$,
suppose vertex $i$ is always unobserved, then we denote by 
$\alpha_t$ the probability that vertex $j$ becomes an unobserved leaf
vertex (i.e., it has no unobserved successor and has only a single
predecessor) at the $t$-th GLR evolution step, and by
$\gamma_{[0,t]}$ the probability that $j$ has been observed at the end of
the $t$-th GLR step. Similarly, suppose the successor vertex $j$ of
a randomly chosen arc $(i, j)$ is always unobserved,
we denote by $\beta_{[0,t]}$ the probability
that the predecessor vertex $i$ has been occupied at the end of the
$t$-th GLR  step, and by $\eta_t$ the probability that at the
end of the $t$-th GLR step vertex $i$ becomes observed but unoccupied and
having no other unoccupied successors except vertex $j$.
These four set of probabilities are related by the following set of
iterative equations:
\begin{subequations}
  \label{eq:abcd}
  \begin{align}
    \alpha_t & = \delta_{t}^{0} Q_{-}(1, 0) +
    \sum\limits_{k_{+},\; k_{-}}
    Q_{-}(k_{+}, k_{-}) \biggl[ \delta_{t}^{1}
      \Bigl[ (\eta_{0})^{k_{+}-1}
        (\gamma_{[0,0]})^{k_{-}} -\delta_{k_{+}}^{1} \delta_{k_{-}}^{0}\Bigr] +
      \nonumber \\
      & \quad 
      (1-\delta_{t}^{0} -\delta_{t}^{1})
      \Bigl[
        \bigl(\sum\limits_{t^\prime=0}^{t-1} \eta_{t^\prime}\bigr)^{k_{+}-1}
        (\gamma_{[0,t-1]})^{k_{-}}
        -\bigl(\sum\limits_{t^\prime=0}^{t-2} \eta_{t^\prime}\bigr)^{k_{+}-1}
        (\gamma_{[0,t-2]})^{k_{-}}
        \Bigr]
      \biggr] \; ,  \nonumber \\
    & \\
    \beta_{[0,t]} & = 1 - \sum\limits_{k_{+}, \; k_{-}}Q_{+}(k_{+}, k_{-})
    \biggl[ \delta_{t}^{0} (1- \delta_{k_{+}}^0)
      (1-\alpha_0)^{k_{-}-1} + \nonumber \\
      & \quad \quad \quad\quad\quad\quad\quad \quad
      (1-\delta_{t}^{0}) 
      \Bigl[ 1- 
        \bigl(\sum\limits_{t^\prime=0}^{t-1} \eta_{t^\prime} \bigr)^{k_{+}} \Bigr]
      (1-\sum\limits_{t^\prime = 0}^{t} \alpha_{t^\prime} )^{k_{-}-1} \biggr]
    \; , \\
    \gamma_{[0,t]} & = 1 - \sum\limits_{k_{+}, \; k_{-}}
    Q_{-}(k_{+}, k_{-}) (1-\beta_{[0,t]})^{k_{+}-1} \bigl(
    1-\sum\limits_{t^\prime=0}^{t} \alpha_{t^\prime}\bigr)^{k_{-}} \; , \\
    \eta_t & = \delta_{t}^0 \sum\limits_{k_{+}, \; k_{-}} Q_{+}(k_{+},
    k_{-}) \bigl(1- (1-\beta_{[0,0]})^{k_{+}}\bigr) (\gamma_{[0,0]})^{k_{-}-1}
    + \nonumber \\
    & \quad \quad
    (1-\delta_{t}^0) 
    \sum\limits_{k_{+}, \; k_{-}} Q_{+}(k_{+}, k_{-}) 
    \Bigl[ \bigl(1- (1-\beta_{[0,t]})^{k_{+}} \bigr)
      (\gamma_{[0,t]})^{k_{-}-1}  \nonumber \\
      & \quad \quad \quad\quad \quad \quad\quad \quad \quad \quad\quad\quad
      -\bigl(1- (1-\beta_{[0,t-1]})^{k_{+}} \bigr)
      (\gamma_{[0,t-1]})^{k_{-}-1}\Bigr] \; .
  \end{align}
\end{subequations}

Let us define $\alpha_{cum}\equiv \sum_{t\geq 0}^{+\infty} \alpha_t$,
$\beta_{cum}\equiv \beta_{[0,\infty]}$,
$\gamma_{cum}\equiv \gamma_{[0,\infty]}$ and
$\eta_{cum}\equiv \sum_{t\geq 0}^{\infty} \eta_{t}$ as the 
cumulative probabilities
over the whole GLR process. From Eq.~(\ref{eq:abcd}) we can
verify that these four cumulative probabilities satisfy the following
self-consistent equations:
\begin{subequations}
  \begin{align}
    \alpha_{cum} & = \sum\limits_{k_{+},\; k_{-}}
    Q_{-}(k_{+},\; k_{-})
    (\eta_{cum})^{k_{+}-1} (\gamma_{cum})^{k_{-}} \; , \\
    \beta_{cum} & = 1- \sum\limits_{k_{+},\; k_{-}}Q_{+}(k_{+},\; k_{-})
    \bigl[1-(\eta_{cum})^{k_{+}} \bigr] (1-\alpha_{cum})^{k_{-}-1} \; , \\
    \gamma_{cum} & = 1-
    \sum\limits_{k_{+},\; k_{-}} Q_{-}(k_{+},\; k_{-}) (1-\beta_{cum})^{k_{+}-1}
    (1-\alpha_{cum})^{k_{-}} \;, \\
    \eta_{cum} &= \sum\limits_{k_{+},\; k_{-}}Q_{+}(k_{+},\; k_{-})
    \bigl[1- (1-\beta_{cum})^{k_{+}}\bigr] (\gamma_{cum})^{k_{-}-1} \; .
  \end{align}
\end{subequations}
The fraction $n_{core}$ of vertices that remain to be unobserved at the
end of the GLR process is 
\begin{eqnarray}
  n_{core}  & = &
  \sum\limits_{k_{+}, \; k_{-}}
  P(k_{+}, k_{-}) \bigl[
    (1-\beta_{cum})^{k_{+}} - (\eta_{cum})^{k_{+}} 
    \bigr] (1-\alpha_{cum})^{k_{-}} \nonumber \\
  & & - \sum\limits_{k_{+}, \; k_{-}}
  P(k_{+}, k_{-}) k_{+} (1-\beta_{cum}-\eta_{cum})
  (\eta_{cum})^{k_{+}-1} (\gamma_{cum})^{k_{-}} \; .
  \label{eq:n}
\end{eqnarray}

The fraction $w$ of vertices that are occupied during the whole
GLR process is evaluated through
\begin{eqnarray}
  w  & = & 1 
  - \sum\limits_{k_{+}, \; k_{-}} P(k_{+}, \; k_{-}) \bigl[1-
    (\eta_{cum})^{k_{+}} \bigr] (1-\alpha_{cum})^{k_{-}} \nonumber \\
  & & - P(1, 0) \eta_{0} 
  -\sum\limits_{t\geq 1} \sum\limits_{k_{+}, \; k_{-}}
  P(k_{+},\; k_{-}) k_{+} \eta_t \bigl(\sum_{t^\prime=0}^{t-1}
  \eta_{t^\prime} \bigr)^{k_{+}-1} 
  \bigl(\sum\limits_{t^\prime=0}^{t-1} \gamma_{t^\prime}\bigr)^{k_{-}}
  \nonumber \\
& & 
  -\sum\limits_{t\geq 1} \sum\limits_{k_{+}, \; k_{-}}
  P(k_{+}, \; k_{-}) k_{-} \alpha_t \bigl(
  \sum\limits_{t^\prime=0}^{t-1} \gamma_{t^\prime} \bigr)^{k_{-}-1}
  \Bigl[1-\bigl(1-\sum\limits_{t^\prime=0}^{t-1} \beta_{t^\prime} \bigr)^{k_{+}}
    \Bigr] \; .
\end{eqnarray}

\end{document}